# An In-depth Analysis of Spam and Spammers

Dhinaharan Nagamalai
*Wireilla Net Solutions Inc,Chennai,India*

Beatrice Cynthia Dhinakaran, Jae Kwang Lee
*Department of Computer Engineering,
Hannam University, South Korea*

## Abstract

*Electronic mail services have become an important source of communication for millions of people all over the world. Due to this tremendous growth, there has been a significant increase in spam traffic. Spam messes up user's inbox, consumes network resources and spread worms and viruses. In this paper we study the characteristics of spam and the technology used by spammers. In order to counter anti spam technology, spammers change their mode of operation, therefore continues evaluation of the characteristics of spam and spammers technology has become mandatory. These evaluations help us to enhance the existing anti spam technology and thereby help us to combat spam effectively. In order to characterize spam, we collected four hundred thousand spam mails from a corporate mail server for a period of 14 months from January 2006 to February 2007. For analysis we classified spam based on attachment and contents. We observed that spammers use software tools to send spam with attachment. The main features of this software are hiding sender's identity, randomly selecting text messages, identifying open relay machines, mass mailing capability and defining spamming duration. Spammers do not use spam software to send spam without attachment. From our study we observed that, four years old heavy users email accounts attract more spam than four years old light users mail accounts. Relatively new email accounts which are 14 months old do not receive spam. But in some special cases like DDoS attacks, we found that new email accounts receive spam and 14 months old heavy users email accounts have attracted more spam than 14 months old light users. We believe that this analysis could be useful to develop more efficient anti spam techniques.*

## 1. Introduction

E-mail has emerged as an important communication source for millions of people world wide by its convenience and cost effectiveness [6]. Email provides user's low cost transfer of messages to large number of people by a click of a button. Email message sizes ranges from 1 kb to several mega bytes which is much larger than fax and other communication devices. The byproducts of email such as instant messaging, chat etc., make life easier and adds more sophisticated facilities to the Internet users. According to a Radicati Group [18] study from the first quarter of 2006, there were about 1.1 billion email users worldwide. Usually the Internet penetration is very high in USA & Europe. But due to the recent upraise of Asian power houses like China and India, the number of email users has increased tremendously over to a million. [15] Due to this increase spam has become a serious threat to the Internet Community [8]. Spam is defined as unsolicited, unwanted mail that endangers the very existence of the e-mail system with massive and uncontrollable amounts of message [4]. Spam brings worms, viruses and unwanted data to the user's mailbox. These spam mails are

sent by spammers. Spammers are well organized business people or organizations that want to make money. DDoS attacks, spy ware installations and worms contribute a large portion to the spam traffic. According to research [5] most spam originates from USA, South Korea, and China respectively. Nearly 80% of all spam are received from mail relays [5].

In this paper we present the characteristics of spam and spammers. We setup a spam trap in our mail server and collected spam for the past 14 months from January 2006 to February 2007 to characterize spam and its senders. Several standard spam tests were conducted to separate spam from the incoming mail traffic. The standard test includes various source filters, content filter. The various source filter tests includes Baysean filter, DNSBL, SURBL, SPF, Grey List, rDNS etc. The learning is enabled in content filters. The size of the dictionary is 50000 words. In our organization we strictly implement mail policies to avoid spam mails. The users are well instructed to use mail service only for communication purposes. From our data we have classified spam into two types. The first type is spam mails with attachment and the second type is spam without attachment. The spam without attachment is further classified as spam containing only text message and with URL or a clickable link. In this paper, we refer URL as links without hyperlinks. The spam with attachments are classified into four categories as spam containing image plus text, image plus URL and text, image plus URL, and spam containing executable files as attachments.

The rest of the paper is organized as follows. Section 2 discusses related work. Section 3 provides data collection of legitimate and spam mails. In section 4, we describe the characteristics of spam traffic. Section 5 provides details of spammers and their technology. Section 6 examines types of email accounts that attract more spam. We conclude in section 7.

## 2. Related work

[1] Presented a multi layer approach to defend DDoS attacks caused by spam. Their study reveals the effectiveness of SURBL, DNSBLs, content filters. They have presented characteristics of virus, worms and trojans accompanied by spam as an attachment. In [3] examines the use of DNS black lists. They have examined seven popular DNSBL and found that 80% of the spam sources are listed in some DNSBL. [4] Presented a comprehensive study of clustering behavior of spammers and group based anti spam strategies. Their study exposed that the spammers have demonstrated clustering structures. They have also proposed a group based anti spam frame work to block organized spammers. [5] Presented a network level behavior of spammers. They have analyzed spammers IP address ranges, modes and characteristics of botnet. Their study reveals that blacklists were remarkably ineffective at detecting spamming relays. Their study states that to trace senders the internet routing structure should be secured. [6] Presented a comprehensive study of spam and spammers technology. His study reveals that few work email accounts suffer from spam than private email. In [8] Gomez, Crsitino presented an extensive study on characteristics of spam traffic in terms of email arrival process, size distribution, the distributions of popularity and temporal locality of email recipients etc., comparison with legitimate mail traffic. Their study shows major differences between spam and non spam mails.

## 3. Data Collection

We characterized spam based on 14 months collection of data over 400,000 spam from a corporate mail server. The mail server provides service to 200 users with 20 group email accounts and 200 individual mail accounts. The speed of the Internet connection is 100 Mpbs for the LAN, with 20 Mbps upload and download speed (Due to security and privacy concerns we unable to disclose the real domain name).To segregate spam from legitimate mails, we conducted standard spam detection tests in our server. The spam mails detected by these techniques were directed to the spam trap that is set up in mail server.

Spammers do not change their tactics on day to day basis. The results of our study show that spammers follow the same technology until the anti spammers find efficient ways to keep them at bay. The time period ranges from 8 months to 1 year. In our collection we see that the spammers follow the same technology from May 2006 to Feb 2007.

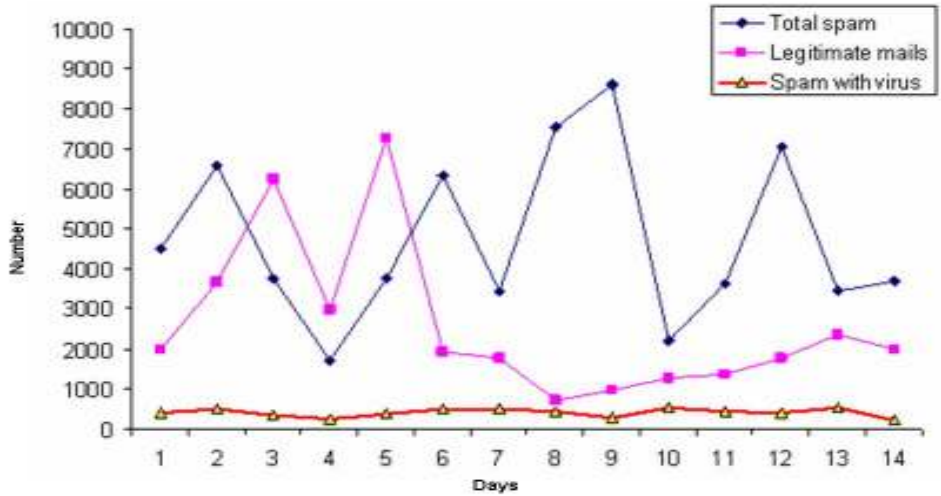

**Figure 1.** Incoming mail traffic

Figure 1 shows the incoming mail traffic of our mail server for 2 weeks. X axis is day and y axis is the number of spam received by end users. It shows number of legitimate mails, spam and spam with virus as an attachment for a period of 2 weeks from February 1 to February 15. The figure shows that the spam traffic is not related to legitimate mail traffic. The legitimate mail traffic is a two way traffic induced by social network [8]. But the spam traffic is one way. Roughly the number of legitimate mails ranges from 720 to 7253 with an average rate of 906 per day. The spam mails ranges from 1701 to 8615 with an average rate of 4736 per day. The spam with viruses as an attachment ranges from 209 to 541 with an average rate of 403 per day. From the graph we can understand that the server handles more number of spam than legitimate mail traffic.

## 4. Spam Classification and Characteristics

We have analyzed millions of spam received in our spam trap. The spam mails are usually related to finance, pharmacy, business promotion, adultery services and viruses. Considerable amount of spam has virus and worms as an attachment [1]. Based on our study the spam mails typically fall into one of two camps, spam without attachment and spam with attachment. The Figure 2 shows the number of spam with attachment and spam without attachment collection for a period of 2 weeks in our corporate mail server. X axis is day and y axis is the number of spam received by our mail server. The mail traffic for spam with attachment ranges from 354 to 4557 with an average rate of 1872 mails per day. The spam mails without attachment received by our spam trap ranges from 346 to 4825 with an average rate of 2722. In terms of traffic volume spam with attachment is not related with spam without attachment. These are generated by different spammers using their own respective technology.

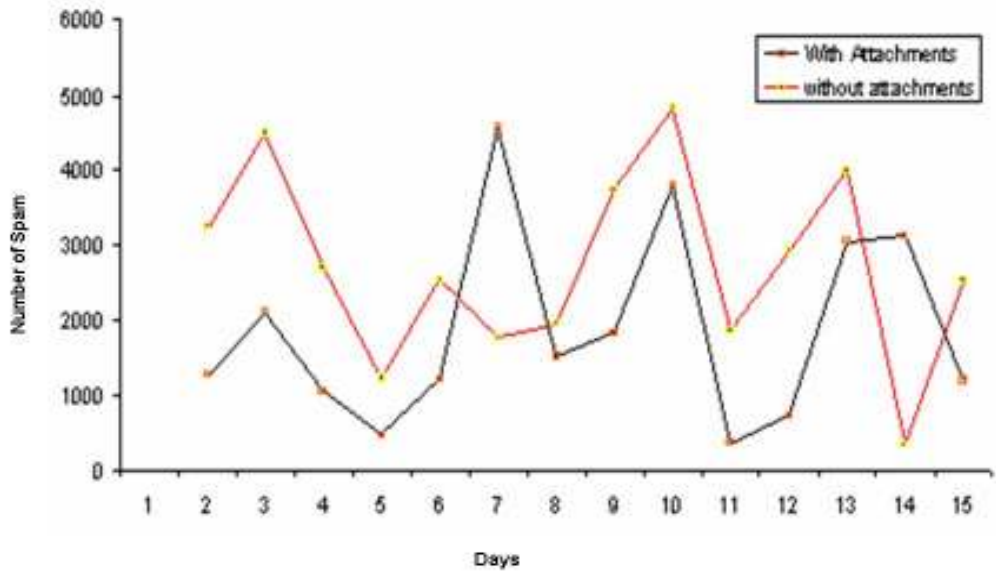

**Figure 2.** Spam with attachment and without attachment traffic

### 4.1 Spam without attachment

Spam without attachments are mostly text messages. It can be roughly categorized as plain text messages and text messages with URL or a clickable link to a website. The size of these message ranges from 2 kb to 3 kb. Spam which contains only text messages are mostly related to scam. Scam mails are sent by individuals to limited number of receivers. Scam related mails are not mass mails, not sent by using specialized software like other spam mails. Most of the sender's mail accounts are fake or don't exist. From our collection we found some messages are sent from Africa using Japanese domains. Spammers have sent these mails to unknown user mail accounts by assumption. Scam mails are less in numbers when compared to other category of spam without attachment. The size of scam mails ranges from 2 to 3 kb.

The spam trap received considerable numbers of spam without attachment. These mails are related to pharmacy. These pharmacy related spam containing text messages and links to their web site. The size of these mails in this category ranges from 1 kb to 3 KB.

**Table 1.** Spam without attachment data collection

| Days | Plain text message | Text + link to the website | Others |
|---|---|---|---|
| 1 | 2271 | 949 | 25 |
| 2 | 1790 | 2682 | 14 |
| 3 | 207 | 2492 | 1 |
| 4 | 489 | 734 | 1 |
| 5 | 169 | 2376 | 1 |
| 6 | 253 | 1520 | 0 |
| 7 | 546 | 2167 | 5 |

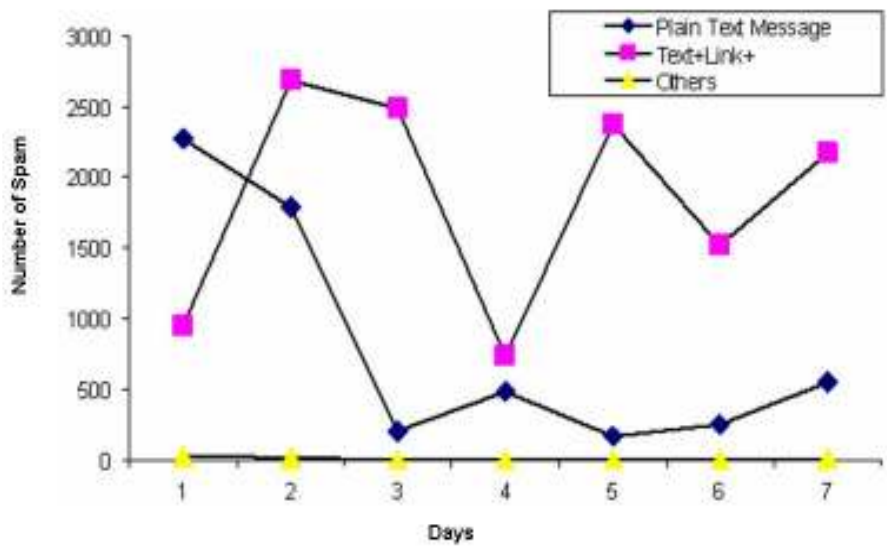

**Figure 3.** Spam without attachment collection

Figure. 3 shows the number of plain text spam mails, spam mails containing text message and a link to the targets website or a URL and different types of spam mails received by our spam trap for a period of seven days. The x axis is day and y axis is the number of spam received in the spam trap. The number of spam containing only text ranges from 169 to 2271 with an average of 818 per day. The number of spam mails with text and a clickable link or an URL ranges from 734 to 2682 with an average rate of 1846. The false positive and some spam error are referred as others in this figure. The others range from 0 to 25 with an average of 8 spam mails. We found that in a week the spam containing text and Clickable link or URL is 100% higher than spam containing only text messages. There is no relation between spam containing plain text and spam with text plus clickable link or URL in terms of traffic volume and traffic pattern.

## 4.2 Spam with attachment

From the spam we have collected, we found that more than 50% of the spam falls under this category. This category of the spam contains a combination of pictures, text and URL or a clickable link to a target web site. The attachments are image files (.gif, .jpg) and executable (.exe) files. The image files are mostly .gif format and rarely .jpg format with size ranging from 5kb to 45 kb. Spam with .exe file as attachments are mail bombs containing viruses and worms [1]. Spam with attachments can be classified into four major categories. The first category of spam mails containing image file (.gif) and text message, the second category of spam containing image, text, URL, the third category of spam containing only image with clickable web link and the fourth category containing worms, virus, Trojans as an attachments. The size of these spam ranges from 13kb to 45 kb with attachments as Image files or executable files. We have monitored spam with attachment traffic for a period of 14 months from Jan 2006 to February 2007. The sample data is illustrated in a table for a period of 7 days.

**Table 2.** Spam with attachment collection for seven days

| Days | Image file +text | Image+Link/URL | Image+Text+URL | Others |
|------|------------------|----------------|----------------|--------|
| 1 | 611 | 505 | 168 | 301 |
| 2 | 1220 | 99 | 188 | 170 |
| 3 | 184 | 121 | 81 | 282 |
| 4 | 17 | 80 | 255 | 85 |
| 5 | 746 | 33 | 195 | 33 |
| 6 | 2005 | 0 | 911 | 1093 |
| 7 | 1820 | 124 | 235 | 156 |

Fig. 4 shows the number of spam with Image file plus text message, image file with clickable link or URL, spam containing text plus image plus URL, and other spam mails received by spam trap for a period of one week. The other spam mail category includes false positive and bulk mail once subscribed by the users. In the figure x axis is day and y axis is the number of spam received in the spam trap. The spam containing Image plus text message ranges from 17 to 2005 with an average of 943 per day. Spam mails containing image and a clickable link or an URL ranges from 0 to 545 with an average rate of 137. The spam containing Image file, text and URL ranges from 81 to 911 with an average of 290 per day. The other spam ranges from 33 to 1093 with an average of 303 spam mails. From the figure, we see that the trap received more number of spam containing Image plus text message than other categories during the week. There is no relation between these categories in terms of spam traffic volume.

The size of the individual spam ranges from 8kb to 40 kb. Our study shows each spam sent by unique ids from unique IP addresses. The spammer used different IP addresses with different mail accounts. These spam were received mostly from spoofed IP addresses with fake mail accounts and by using relays [9] [5]. Subscribed mails like news letters and commercials like telecom and home appliances related are also considered as a spam by most users. These types of commercials are easy to unsubscribe and it is negligible in the total spam percentage. The subscribed mails contain pictures, text, and animation as contents with more than 50 kb as their size.

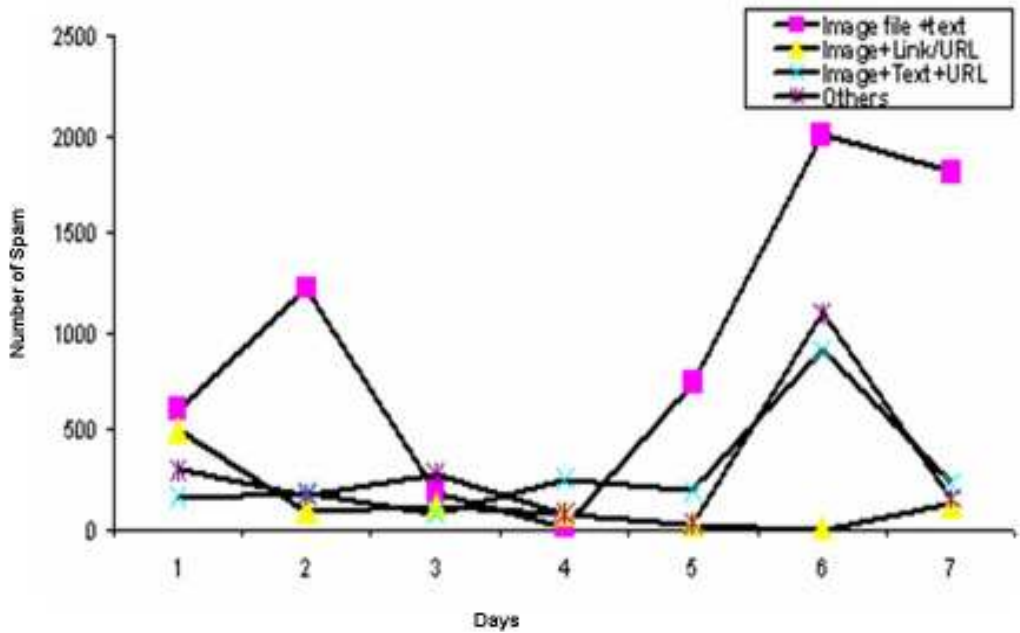

**Figure 4.** Spam with attachment collection

Spam containing text, image, URL: In this category of spam contains text, image and URL. The text is unrelated to the intended spammers business. The image contains URL which is directly linked to the intended business organization. The size of the spam ranges from 10 kb to 30 kb. A considerable number of spam of this type are responsible for phishing attacks.

Spam containing image and text: In this category we discuss spam containing text and image as a message. The image contains text and URL of online business. Usually this type of spam doesn't have clickable links. The contents of the image are relevant to the intended business organization. But the text following the images are randomly selected by spam software and are not related to the intended message. The size of text message ranges from a single line to several paragraphs. By changing the contents of text and its size, the spammers try to confuse the filters. In this category we found that the majority of spam are related to pharmacy and finance. Our study shows in this category more than 50% of the spam are related to financial matters. The size, color and contents of the image are same for a particular period of time. The spammer's main intention here is to deliver a message available inside an image. To confuse the filters, the spammers add different size of text message.

Spam containing image and URL or a clickable link: A considerable number of spam contains image file plus URL or a clickable link. Since the URL or a clickable link is placed inside the image, thus type of spam can easily bypass content filters. The fourth category of spam contains .exe (executable) file as an attachment. The executable files are mostly virus, worms, trojon etc,. The attachment sizes ranges from 35 k to 140 kb. These mails are intended to spread virus, try to establish mail bombs to mount DDoS attack to the server and the network. If the attachment is executed, it will drop new files in windows folder and change the registry file contents, link to the attacker's website to download more programs to harm the network further. The infected machine collects email addresses from the windows address book and automatically sends mails to others in the same domain [1].

Distributed Denial of Service (DDoS) attack is a large scale, coordinated attack on the availability of services at a victim system or network resource [5]. DDoS attack through spam mail is one of the newest versions of common DDoS attack. In this type, the attacker penetrates the network by a small program attached to the spam mail. After the execution of the attached file, the mail server resources will be eaten up by mass mails from other machines in the domain thus resulting denial of services. The spam contains small size of .exe file as an attachment (for example update.exe). The attackers used double file extension to confuse the filters (Update_KB2546_*86.BAK.exe (140k)) and user. The attachment size ranges from 35 to 180 KB. The names of the worms used in these kind of DDoS attacks are WORM_start.Bt, WORM_STRAT.BG, WORM_STRAT.BR, TROJ_PDROPPER.Q. Upon execution, these worms drop files namely serv.exe, serv.dll, serv.s, serv.wax, E1.dll, rasaw32t.dll etc. DDoS malware cause direct and indirect damage by flooding specific targets [16]. Mass mailers and network worms cause indirect damage when they clog mail servers and network bandwidth. It will also consume the network bandwidth and resources, causing slow mail delivery and further resulting Denial of service. The server slow down due to enormous request from clients and bulk mail processing [1]. We have analyzed the spam trap to measure the traffic of spam with virus, worm and trojon.

**Table 3.** Spam with virus, worms and trojons traffic for 7 weeks and 7 days of the first week

| Weeks | Number of Spam with virus, worms, trojons | Days of first week | Number of Spam with virus, worms, trojons |
|---|---|---|---|
| 1 | 2847 | 1 | 407 |
| 2 | 1947 | 2 | 485 |
| 3 | 1152 | 3 | 339 |
| 4 | 1834 | 4 | 253 |
| 5 | 2423 | 5 | 384 |
| 6 | 1245 | 6 | 494 |
| 7 | 2256 | 7 | 486 |

Figure. 5 shows the number of spam containing virus, worms and trojon as an attachment for a period of seven weeks. The x axis is week and y axis is the number of spam with virus, worms and trojon as an attachment, received in our server spam trap. These types of spam received ranges from 1245 to 2847 with an average of 1957 per week. We further analyzed these spam for a week to present traffic for a day. Figure 6 shows the number of spam with virus received in our mail server for seven days. The spam mails with virus, worms and trojons ranges from 253 to 496 with an average rate of 407. The x axis is marked day and y axis is the number of spam with virus, worms and trojon as an attachment, received in our server spam trap. There is no relation between spam with virus traffic and legitimate mail traffic.

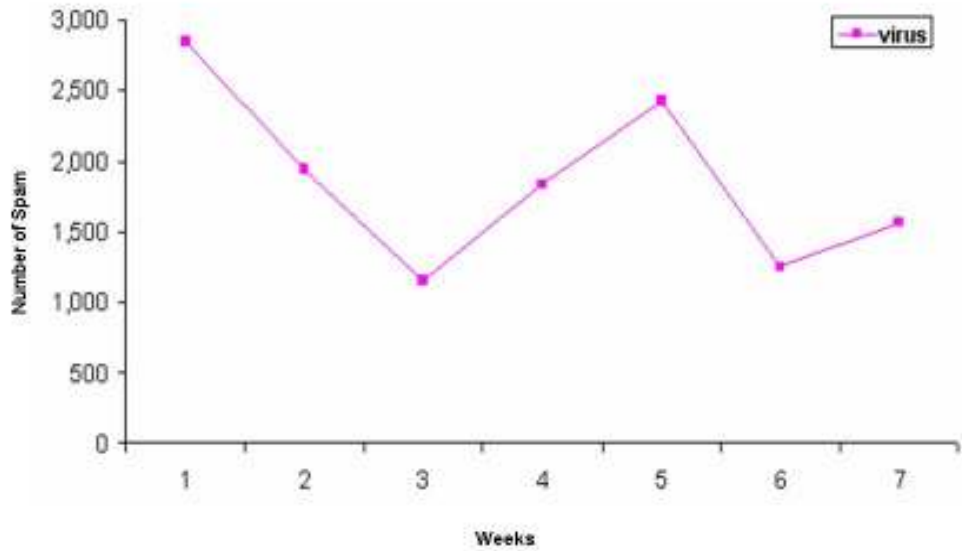

**Figure 5.** Spam with virus, worms and Trojons traffic for 7 weeks

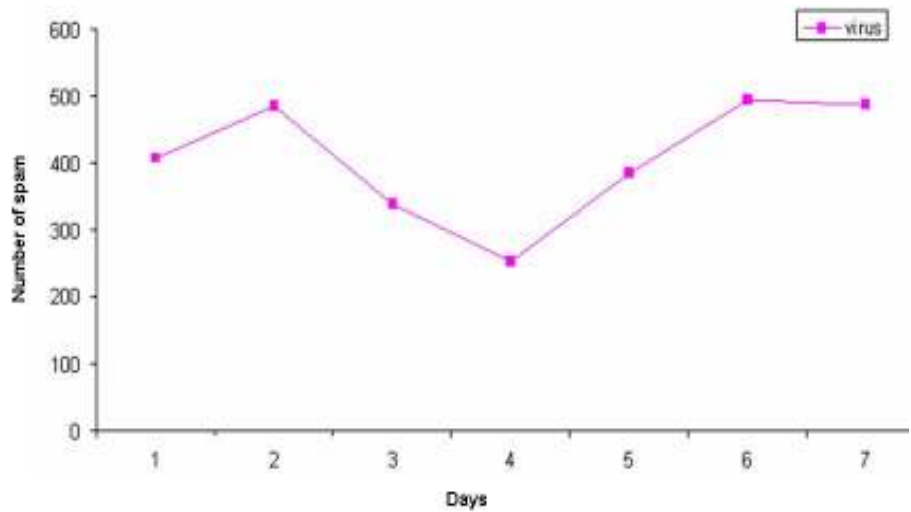

**Figure 6.** Spam with virus, worms and Trojons traffic in the first week of figure 5

From our analysis we have identified a list of frequently spammed virus, worms and trojons for a period of a month from February 1 to February 28. The list is shown in table 4.

**Table 4.** Top 10 Virus, worms, trojons received by spam trap on February 2007

| Virus,worms,trojon | Number of spam mails |
|---|---|
| Email-Worm.Win32.NetSky.q | 630 |
| Spam.Phish.url | 3278 |
| Email-Worm.Win32.Bagle.gt | 556 |
| Spam.Porn.PORN_NB_PORNHINT_1 | 231 |
| NBH-BHIDIFM | 118 |
| Net-Worm.Win32.Mytob.c | 150 |
| Email-Worm.Win32.Zhelatin.o | 110 |
| Email-Worm.Win32.Zhelatin.u | 58 |
| Email-Worm.Win32.Bagle.gen | 97 |
| Trojan-Downloader.Win32.Agent.bet | 161 |
| Trojan-Downloader.Win32.Small.dam | 85 |

## 5 Spammers Technology

### 5.1 Spam without attachment

In case of text messages with link, we have noticed 2 categories of senders. In the first category we found spammers use free mail service providers like yahoo and other unfamiliar mail services. In this case it is a single person sends spam to many users with multiple fake accounts with the same contents. Mostly the number of receivers are 2, very rarely it exceeds more than 2, and if it does it usually less than 10. The destination mail accounts are normally collected from other spammers or by assumption. The contents of spam include pharmacy, adult services, finance etc. The links available are legitimate websites or blogs, not directly related to the spammer's website. From these legitimate website or blog, the user will be redirected to the spammer's website. The legitimate website is used to confuse the anti-spam techniques. We have observed that spammers use Yahoo, geocities as their intermediate machine to redirect their link as shown in the example below.

http://us.rd.yahoo.com/*http://jpsvpd.miserpoison.com/legalrx/?38947779
http://www.geocities.com/hi_bumukybew/ redirected to http://www.enjoycasualsex.com/

Our study shows that mostly this category of spam mail originates from the US. The second category of spammers own small sized web spaces with limited mail accounts. The website has unwanted data or "under construction" notice. Most spammers use European (non English speaking) or Asia Pacific region ISPs as a host. The contents of these spam are related to casinos and pharmacy. Plain text messages are very small in number when compared to other kind of spam. Example for this kind of mail is the Nigerian Fraud Email, etc, [15][16]

### 5.2 Spam with attachment

The senders of spam with attachment use highly sophisticated software tools. The spam is sent by using well designed software or a small program. From our analysis we have identified that 100% of the sender's mail accounts don't exist i.e. fake mail accounts. Spammers don't have their own website or domain. The spam header shows only one receiver in the mail but that is not true. These spam arrive from a relay machine which makes it difficult to find the real sender. Spammers use spoofed ip address to hide their identity.

There are many software available to make spamming more sophisticated and easy [1]. For example Phasma Email Spoofer, Bulk Mailer, Aneima 2.0, Avalanche 3.5, Euthanasia etc. Spam software use random servers to create fake mail accounts and the spammer can fix the number of mails to send and the period of the attack. It does facilitate the spammer to use relay machines. By using this software, spammers can send mass mails to end users without showing the destination mail accounts. The software get guidelines from the spammer to create a header, otherwise it will generate its own fake header. The spammer can attach their intended file as an attachment with random or fixed text messages. There are software which can send 365 spam per minute. The mail sent by using sophisticated software pretends to be sent by Microsoft outlook express [12]. Since the messages contain only HTML content, these messages were not sent by using Microsoft outlook express. But the spam message contain virus, worms, trojan as an attachment doesn't show that they were sent by Microsoft outlook express. Regardless of spam type, mostly spam sent with the help of software indicates that which are sent by Microsoft outlook express.

## 6. Spam vulnerable email accounts

There are many ways to get end users e-mail accounts. Many professional spammers sell e-mail accounts to other spammers for cheap prices. The professional email account sellers collect e-mail accounts by using well designed software and also offer free bees. We are not going to discuss this, instead we are going to discuss about email accounts which attract more spam. Heavy users receive more spam than others [6]. Heavy users can be defined as those who have their own web site, blogs and those who actively involve in forums and chat rooms. Apart from spammers, legitimate business houses like Internet, telecom service providers also send spam to unknown end users. Various organizations including research organizations, job fair organizers, religious organizations, also send spam to unknown users. The organizers get thousands of end user's mail accounts from other organizers to spam end users. Most organizers don't provide the facility to unsubscribe from their mailing list.

Regardless of end users contacts and activities, email accounts created several years ago receive more spam than others. To analyze this, we have selected 20 email accounts from our network and monitored the spam traffic. Out of 20, ten email accounts are four years old, in which five accounts are heavy users and remaining are non heavy user accounts. The heavy users are those who have their maintain own website, blogs and actively involve in various forums and chat rooms [6]. The selected heavy users email accounts were published in various newspaper advertisements and others. The remaining 10 email accounts are new email accounts created 14 months ago. In this category, 5 mail accounts belong to heavy users; the remaining 5 accounts are non heavy user accounts. We have monitored the spam traffic received by these 20 accounts for a period of 6 weeks from Jan 1, 2007 to Feb 15, 2007. The spam traffic is illustrated in the table.5.

When we say old email account it means accounts that were created four years ago. From our study, we had found out that old heavy user email accounts attracted more spam than others. The data collection for six weeks is illustrated in figure 7. The x axis represents weeks from January 1 to February 15. The Y axis is the number of spam received by end users. From the graph, we see that the four years old heavy user email accounts received spam ranging from 78 to 1015, which is an average of 532 per week. They attracted roughly 45% more spam than four years old light users mail accounts. Due to DDoS attack in the network we see more spam has been received in the fourth week.

**Table 5.** Number of spam received by all types of email accounts

| User | Number of Spam received | | | | | |
|---|---|---|---|---|---|---|
| | Week 1 | Week 2 | Week 3 | Week 4 | Week 5 | Week 6 |
| Old Heavy user 1 | 78 | 756 | 408 | 1025 | 98 | 665 |
| Old Heavy user 2 | 299 | 410 | 544 | 877 | 310 | 590 |
| Old Heavy user 3 | 252 | 1015 | 862 | 1045 | 142 | 890 |
| Old Heavy user 4 | 157 | 864 | 589 | 920 | 189 | 601 |
| Old Heavy user 5 | 283 | 929 | 544 | 1200 | 320 | 989 |
| Average | 213.8 | 794.8 | 589 | 1013 | 211 | 747 |
| Old Light User 1 | 456 | 302 | 952 | 1089 | 478 | 319 |
| Old Light User 2 | 377 | 194 | 589 | 640 | 345 | 225 |
| Old Light User 3 | 550 | 151 | 136 | 768 | 490 | 187 |
| Old Light User 4 | 141 | 863 | 589 | 987 | 167 | 641 |
| Old Light User 5 | 141 | 108 | 45 | 234 | 128 | 120 |
| Average | 333 | 323 | 462 | 892 | 321 | 298 |
| New Heavy user 1 | 0 | 0 | 0 | 15 | 0 | 0 |
| New Heavy user 2 | 0 | 0 | 0 | 8 | 0 | 0 |
| New Heavy user 3 | 0 | 0 | 0 | 10 | 0 | 0 |
| New Heavy user 4 | 0 | 0 | 0 | 14 | 0 | 0 |
| New Heavy user 5 | 0 | 0 | 0 | 12 | 0 | 0 |
| Average | 0 | 0 | 0 | 11 | | |
| New Light User 1 | 0 | 0 | 0 | 9 | 0 | 0 |
| New Light User 2 | | | | 8 | 0 | 0 |
| New Light User 3 | 0 | 0 | 0 | 9 | 0 | 0 |
| New Light User 4 | 0 | 0 | 0 | 7 | 0 | 0 |
| New Light User 5 | 0 | 0 | 0 | 2 | 0 | 0 |
| Average | 0 | 0 | 0 | 7 | 0 | 0 |

When we say old email account it means accounts that were created four years ago. From our study, we had found out that old heavy user email accounts attracted more spam than others. The data collection for six weeks is illustrated in figure 7. The x axis represents weeks from January 1 to February 15. The Y axis is the number of spam received by end users. From the graph, we see that the four years old heavy user email accounts received spam ranging from 78 to 1015, which is an average of 532 per week. They attracted roughly 45% more spam than four years old light users mail accounts. Due to DDoS attack in the network we see more spam has been received in the fourth week.

The four year old light user (non heavy users) accounts received more spam similar to heavy user's accounts, ranges from 45 to 952 averages 372. The 4 years old non heavy user mail accounts attracted less spam than 4 years old heavy user mail accounts as shown in figure. Regardless whether heavy or light user, all old email accounts attracted more spam than 14 months old new email accounts.

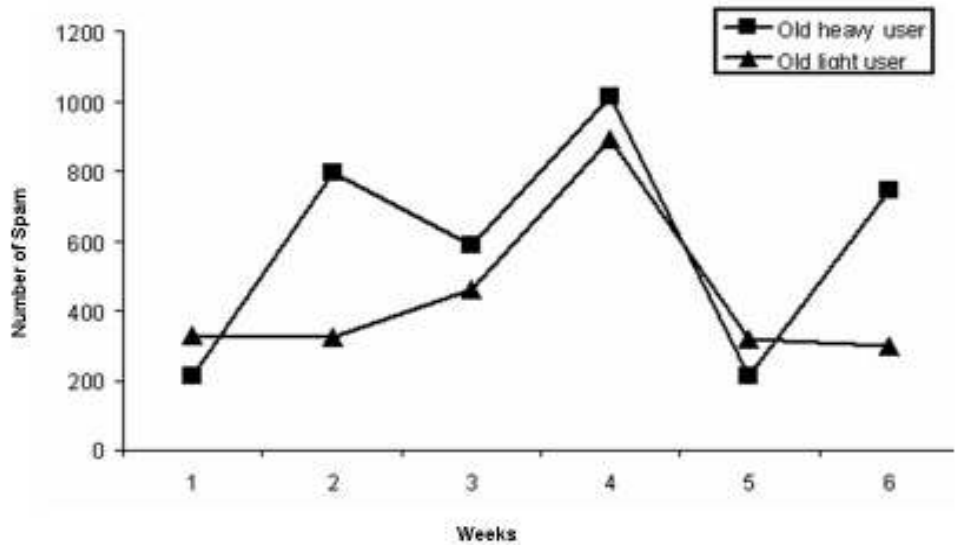

**Figure 7.** Average number of spam received by old heavy users and old non heavy users accounts for 6 weeks.

In figure 8, the x axis is week and y axis is the number of spam received by end user accounts. Roughly the spam ranges from 0 to 11 with an average rate of 2. The 14 months old relatively new accounts didnot attract spam except during DDoS attack period. Since the network suffered by DDoS attack in fourth week, it has received more spam as shown in figure. The 14 months old heavy user email accounts didn't receive spam same as the 14 months old light user mail accounts. Regardless of the nature of the account, 14 month old new mail accounts didnot receive spam. Literally the new email accounts were free from spam for particular period of time, in our case it was up to 14 months. As shown in figure 8, 14 months old heavy user account received more spam than 14 months old light user accounts.

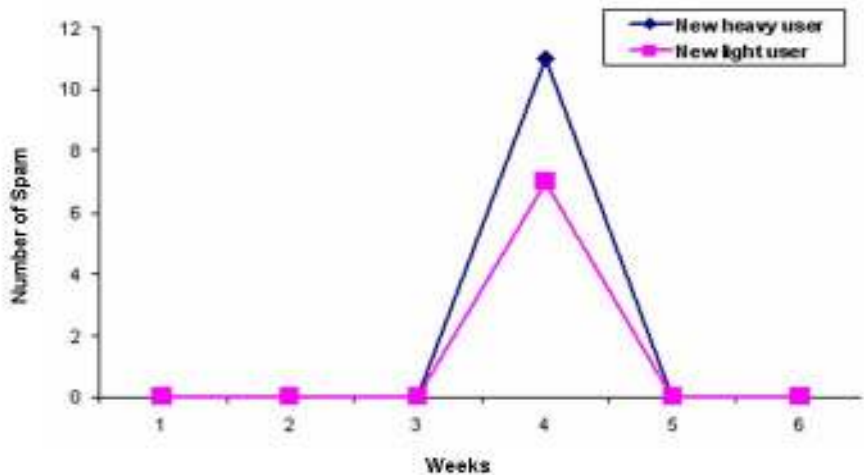

**Figure 8.** Average number of spam received by new heavy users and new non heavy users accounts for 6 weeks from January 1 to February 15.

## 7. Conclusion

From our study we conclude that spam can be classified into 2 major categories. The first category is spam with attachment and the second category is spam without attachments. Spam without attachment are text messages and links to the intended target. Spam with attachments can be classified into 4 types such as spam mails containing image file and text message, spam containing image plus text plus URL, spam containing only image plus clickable web link and spam containing worms, virus, and trojans as attachments. Spam without attachments are small in size, when compared to spam with attachments. The volume of spam traffic is not related to legitimate mail traffic. From our study we understood the technology behind spammers. For spam without attachment, senders use non sophisticated methods. But for spam with attachment, senders use sophisticated software to spam end users. The spamming software provides various facilities like multi targets, spoofing identity, using relays etc., Due to this, tracing senders is difficult.   We have also analyzed the types of email accounts that attract more spam. From our study we conclude that old heavy user email accounts receive more spam than relatively new email accounts. The four years old heavy user email accounts received 45% more spam than four years old light user accounts. The 14 months old relatively new accounts didnot attract spam except DDoS attack period.